\documentclass[runningheads]{svmult}

\usepackage{makeidx}   % allows index generation
\usepackage{graphicx}  % standard LaTeX graphics tool
\usepackage{subeqnar}  % subnumbers individual equations
\usepackage{multicol}  % used for the two-column index
\usepackage{physprbb}  % modified textarea for proceedings,
\makeindex             % used for the subject index
\begin{document}
\title*{The Genesis of Super Star Clusters\protect\newline as Self-Gravitating HII Regions}
\toctitle{The Genesis of Super Star Clusters
\protect\newline as Self-Gravitating HII Regions}
% allows explicit linebreak for the table of content
%
%
\titlerunning{Genesis of Super Star Clusters}
% allows abbreviation of title, if the full title is too long
% to fit in the running head
%
\author{Jonathan C. Tan\inst{1}
\and Christopher F. McKee\inst{2}}
\authorrunning{Tan \& McKee}
% if there are more than two authors,
% please abbreviate author list for running head
%
%
\institute{Dept. of Astronomy, UC Berkeley, Berkeley, CA 94720, USA
\and Depts. of Physics and of Astronomy, UC Berkeley, Berkeley, CA 94720, USA}

\maketitle              % typesets the title of the contribution

\begin{abstract}
We examine the effects of ionization, radiation pressure and main
sequence winds from massive stars on self-gravitating, clumpy
molecular clouds, thereby modeling the formation and pre-supernova
feedback of massive star clusters. We find the process of ``turbulent
mass loading'' is effective in confining HII regions. Extrapolating
typical Galactic high-mass star forming regions to greater initial gas
cloud masses and assuming steady star formation rates, we determine
the timescales for cloud disruption. We find that a dense ($n_c\simeq
2\times 10^5\:{\rm cm^{-3}}$) cloud with initial mass $M_{\rm
c}\simeq 4\times 10^5\:{\rm M_\odot}$ is able to form $\sim 2\times
10^5\:{\rm M_\odot}$ of stars (50\% efficiency) before feedback
disperses the gas after $\sim 3$ Myr. This mass and age are typical of
young, optically visible super star clusters (SSCs). The high
efficiency permits the creation of a bound stellar system. 
%Our
%feedback model provides a physical mechanism for understanding the
%differences between Galactic and super star clusters.
\end{abstract}

\section{Introduction}
Most Galactic stars are born in highly clustered regions
\cite{lad93,car00}, where the disruptive effects of massive stars are
paramount. The multitude of dusty high redshift sources and the
intensity of the far infrared background they produce also imply that
a major fraction ($\sim 1/2$) of total cosmic star formation has
occurred in starbursts, replete with massive stars \cite{tan99}. A
significant fraction of star formation in local starbursts occurs via the
creation of super star clusters (SSCs) \cite{whi99,zha99}, each with
hundreds to thousands of OB stars crammed into a few parsecs. At least
some SSCs are gravitationally bound \cite{ho96a,ho96b} and their
masses ($\sim 10^5 - 10^6\:{\rm M_\odot}$ \cite{ste98,men00}) and
sizes suggest we may have found globular clusters in their infancy.

Massive stars violently disrupt their surroundings with ionizing and
non-ionizing photons, protostellar, main sequence and post-main
sequence winds, and supernovae. Adding to this complexity is the
extremely dense, clumpy and turbulent nature of the gas in which
high-mass stars are born. We present a simplified model of pre-supernova
feedback, to examine how the efficient star formation required to
produce bound clusters may occur in the presence of vigorous energy
injection from many massive stars.

\section{Initial Conditions -- a Clumpy Molecular Cloud}

We consider a spherical cloud of radius, $R_{\rm c}$, mass $M_{\rm
c}$, and mean H density $n_{\rm c}$. The cloud consists of dense
clumps embedded in a uniform inter-clump medium. The clumps are
distributed uniformly within a central core of radius 0.2~$R_{\rm c}$,
and with a $r^{-1}$ distribution outside, which mimics observed
molecular cloud profiles. We choose mean clump mass and volume
fractions of $f_m=0.8$ and $f_V=0.03$ respectively. Our clumps, of
uniform density $n_{\rm cl}$, have a mass spectrum ${\rm d\: {\cal
N}_{\rm cl}}/{\rm d\:ln}\:m_{\rm cl} \propto m_{\rm cl}^{-0.6}$ between upper,
$m_{\rm cl,u}$, and lower, $m_{\rm cl,l}$, limits. Clump masses span a
range of $10^3$, with $m_{\rm cl,u}=0.025 M_{\rm c}$, so no one clump
dominates the cloud. The clump velocities are set so the cloud
is in virial equilibrium.

Plume et al. \cite{plu97} determined sizes and masses for 25 regions
of Galactic high-mass star formation. The mean properties of this
sample were $R_{\rm c} \sim 0.5\:{\rm pc}$, $M_{\rm c}\sim 3800\:{\rm
M_{\odot}}$ (virial mass) and thus $n_{\rm c}\simeq 2\times 10^5\:{\rm
cm^{-3}}$ ($\simeq3.2\times 10^4\:{\rm M_{\odot}\:pc^{-3}}$). Note that
these regions are only a small fraction of the host Giant Molecular
Cloud. Our adopted clump mass and volume fractions imply clump
densities of $n_{\rm cl}=6.5\times 10^6\:{\rm cm^{-3}}$ and an inter-clump density of
$5.0\times 10^4\:{\rm cm^{-3}}$.
%The mean velocity dispersion was
%$\sigma_{{\rm 1D}}=2.6\:{\rm km\:s^{-1}}$. 
These properties form the basis of our model {\bf A}. Determining the
gas properties of the precursors to SSCs is more difficult as current
observations only probe scales down to $\sim~30-100\:{\rm pc}$
\cite{gil00,wil00}, while the star clusters have a typical radius
$\sim 4$~pc \cite{whi99}. For simplicity we consider models with 10
({\bf B}) and 100 ({\bf C}) times the mass of {\bf A}, but with the
same mean density. With the same values of $f_V$ and $f_m$ the clump
and interclump densities are also the same. For each cloud we shall
consider a fixed star formation rate, $\phi_{50}$, which converts 50\%
of the initial cloud into stars over 3~Myr -- i.e., the approximate
time before the first supernova is expected to occur.\footnote{This rate
excludes the formation of an initial star cluster of $\sim250\:{\rm
M_\odot}$ necessary to produce an ionizing luminosity of $10^{49}$
ionizing photons per second (see below).} The properties
of our initial gas clouds are listed in table \ref{Tab1}.

\begin{table}
\caption{Model Parameters}
\begin{center}
\renewcommand{\arraystretch}{1.4}
\setlength\tabcolsep{5pt}
\begin{tabular}{ccccc}
\hline\noalign{\smallskip}
Model & $M_{\rm c}\:{\rm /(M_{\odot})}$ & $R_{\rm c}\:{\rm /(pc)}$ & $\sigma_{\rm 1D}\:{\rm /(km\:s^{-1})}$ & $\phi_{\rm 50}\:{\rm /(M_{\odot}\:yr^{-1})}$\\
\noalign{\smallskip}
\hline
\noalign{\smallskip}
A & $4\times 10^3$ & 0.5 & 2.8 & $5.8\times 10^{-4}$\\
B & $4\times 10^4$ & 1.1 & 5.9 & $6.6\times 10^{-3}$\\
C & $4\times 10^5$ & 2.3 & 12.9 & $6.7\times 10^{-2}$\\
\hline
\end{tabular}
\end{center}
\label{Tab1}
\end{table}

\section{Feedback Processes\label{S:feedback}}

To model the feedback processes, we need the total number of hydrogen
ionizing ($\lambda<912\:{\rm \AA}$) photons emitted per second ($S$),
the bolometric luminosity ($L_{\rm bol}$) and the mass flux
($\dot{M}_w$) and velocity ($v_w$) of the stellar wind from the
forming star cluster.  We derive these quantities from the STARBURST99
model \cite{lei99} at solar metallicity. As we are interested only in
the first few Myr of stellar evolution, we approximate the values of
these quantities to be constant for a given stellar mass. Motivated by
observations of the stellar population of the R136 cluster in 30
Doradus \cite{bra00}, we consider a stellar initial mass function
(IMF) represented by a broken power law, ${\rm d} {\cal N}_*/ {\rm
d}\:{\rm ln}\: m_* \propto (m_*/ m_{*,0})^{-\alpha}$,
%\begin{equation}
%\label{imf}
%\frac{d \eta}{d\:{\rm ln} m} \propto \left(\frac{m}{m_0}\right)^{-\alpha},
%\end{equation}
with $\alpha=1.35$ (Salpeter) for $m_*>m_{*,0}=3\:{\rm M_\odot}$ and
$\alpha=0.8$ for $m_*<m_{*,0}$. We choose lower and upper limits of
stellar mass of $0.1\:{\rm M_\odot}$ and $100\:{\rm M_\odot}$,
respectively. This increases the mass per unit of feedback (assuming
negligible contribution from stars below $m_{*,0}$) by a factor of
1.35 compared to the standard STARBURST99 IMF ($\alpha=1.35$ between
$1\:{\rm M_\odot}$ and $100\:{\rm M_\odot}$). We find
\begin{equation}
\label{s_cluster}
S_{49}=3.94\times 10^{3} M_{*,6},
\end{equation}
\begin{equation}
\label{lbol_cluster}
L_{\rm bol,5}=1.10\times 10^4 M_{*,6},
\end{equation}
\begin{equation}
\label{mdot_cluster}
\dot{M}_{w,-6}=3.09\times 10^3 M_{*,6},
\end{equation}
\begin{equation}
\label{mom_cluster}
%\dot{M}v_w=5.01\times 10^{25} M_*^\prime \:{\rm g\:cm\:s^{-2}}.
\dot{M}_{w,-6} v_{2000}=3.98 \times 10^3 M_{*,6},
\end{equation}
where $S_{49}=S/10^{49}\:{\rm photons\:s^{-1}}$, $L_{\rm bol,5}=L_{\rm
bol}/10^5\:{\rm L_\odot}$, $\dot{M}_{w,-6}=\dot{M}_w/10^{-6}\:{\rm
M_\odot\:yr^{-1}}$, $v_{w,2000}=v_w/2000\:{\rm km\:s^{-1}}$ and
$M_{*,6}=M_*/10^6\:{\rm M_\odot}$. The number of $m_*>8\:{\rm
M_\odot}$ stars (i.e. number of core collapse supernovae) is ${\cal
N}_{\rm SN}=1.4\times 10^4 M_{*,6}$.

\subsection{Ionization}

Assuming spherical symmetry with the massive stars forming at the
center of the cloud, we calculate the ionizing flux received at a
given distance $R$ from the star cluster by accounting for attenuation
by clump shadowing, H recombinations and dust absorption, so that
\begin{equation}
\label{atten}
\frac{dS}{dR}=-S \bar{A}_{\rm cl} {\cal N}_{\rm cl} -4\pi R^2 \alpha_2 n_i n_{\rm e} - \sigma_d n_i S,
\end{equation}
where $\bar{A}_{\rm cl}$ is the mean cross sectional area of clumps at
$R$, ${\cal N}_{\rm cl}$ is the space number density of clumps,
$\alpha_2$ is the recombination coefficient,
% $\alpha_2=2.6 \times
%10^{-13} \:{\rm cm^3\:s^{-1}}$ is the recombination coefficient,
$n_i$ 
%and $n_{\rm e}=1.1n_{\rm H}$ 
is the hydrogen nuclei 
%and electron
number density of the ionized gas and $\sigma_d=0.5\times
10^{-21}\:{\rm cm^2}$ \cite{bal91} is the dust absorption cross-section
per H nucleus. We assume He is single ionized and $n_{\rm He}=0.1
n_i$. An HII region rapidly forms with typical size $R_{\rm
St}\simeq 3.05\times 10^{-2} S_{49}^{1/3} n_{i,5}^{-2/3}\:{\rm
pc}$, for dust free, uniform density gas. The short sound crossing
time justifies our assumption of uniform $n_i$. Thermal balance maintains a
constant ionized gas temperature, $T_i\simeq 10^4\:{\rm
K}$. Over-pressurized compared to the surrounding neutral medium, the
HII region tends to expand.\footnote{For the propagation of ionization fronts into the clump and interclump material, we assume the neutral gas has a temperature of 80 K.}
%, where
%$n_i$ is the H nuclei number density, He is singly ionized and $n_{\rm
%He}=0.1 n_{i}$. 
However, when neutral clumps become exposed to ionizing photons, they
implode and inject mass into the HII region \cite{ber89}. A
compressed neutral globule remains, which continues to evaporate more
slowly \cite{ber90}. We employ models for magnetically supported
clumps so that the pressure of the ionized gas streaming from the surface of an
imploded clump at distance $R$ is
\begin{eqnarray}
\frac{p_{\rm c}}{k} & \simeq & 8.7\times 10^7 n_{\rm cl,7}^{1/21} \left(\frac{S_{49}}{R_{\rm pc}^2}\right)^{4/7} m_{\rm cl}^{-4/21}\:{\rm K\:cm^{-3}}\nonumber\\ 
 & = & 9.9\times 10^9 n_{\rm cl,7}^{1/21} \left(\frac{M_{*,6}}{R_{\rm pc}^2}\right)^{4/7} m_{\rm cl}^{-4/21}\:{\rm K\:cm^{-3}},
\end{eqnarray}
where $m_{\rm cl}$ is the clump mass in solar mass units and
$n_{\rm cl}$ is the initial clump density. Clumps may only
photoevaporate if this pressure is greater than the wind ram pressure
(below). Evaporating clumps are gradually ejected from the HII region
by the rocket effect. However, clumps orbiting in the cloud potential
continue to enter the HII region and inject fresh gas. The HII region
is limited by a recombination front as the ionized gas flows out at
approximately its sound speed. We have examined the details of this
mechanism of ``turbulent mass loading'' and its implications for
Galactic ultracompact HII regions elsewhere \cite{tan01}.

\subsection{Winds}

Stellar winds carve out a hot, low density cavity at the center of the
HII region. The wind's thermal pressure is suppressed by cooling
caused by mass injected by photoevaporating clumps
\cite{mck84}. Instead, we calculate the edge of the wind cavity by
balancing the thermal pressure of the HII region with the wind ram
pressure,
\begin{equation}
\frac{p_{\rm w,ram}}{k}=7.6\times 10^{5}\frac{\dot{M}_{-6} v_{2000}}{R_{\rm pc}^2}\:{\rm K\:cm^{-3}}\simeq 3.0\times 10^9  \frac{M_{*,6}}{R_{\rm pc}^2}\:{\rm K\:cm^{-3}}.
\end{equation}
In addition, this pressure contributes to the clump dynamics and may
quench photoevaporation, particularly at small distances from very
massive clusters. We treat photoevaporation mass injection into the wind via a
supersonic mass-loaded wind model, e.g. \cite{dys95}. In cases of
extreme mass loading the ionizing photons may be trapped in the wind.

\subsection{Radiation Pressure}
Our fiducial clumps absorb most of the momentum flux of the star cluster's
radiation. The pressure is
\begin{equation}
\label{prad_cluster}
\frac{p_{\rm rad}}{k}=8.7\times 10^{9}\frac{M_{*,6}}{R_{\rm pc}^2}\:{\rm K\:cm^{-3}}.
\end{equation}

%(recoil from photoemitted electrons may also be
%important; Weingartner \& Draine 1999). From Vacca, Garmany \& Shull
%(1996) we have for single stars $p_{\rm
%rad}/k~=~2.18~\times~10^{7}~S_{49}^{3/4}~/~R_{\rm pc}^2$, accurate to
%$\pm 10\%$ for 03-09 stars. Thus for our star cluster
%\begin{displaymath}
%\frac{p_{\rm rad}}{k}=2.8\times 10^{9}\frac{M_{*,3}^\prime}{R_{\rm pc}^2}.
%\end{displaymath}
%Radiation pressure typically dominates over $p_{\rm c}$, particularly
%closer to the cluster.

%The most extreme case observed by Plume et al was W51
%M, with $R_{\rm c} \sim 0.6\:{\rm pc}$, $M_{\rm c}\sim 1.6\times
%10^4\:{\rm M_{\odot}}$, $n_{\rm c}\simeq 5.3\times 10^5\:{\rm
%cm^{-3}}$ and $\sigma_{{\rm 1D}}=4.8\:{\rm km\:s^{-1}}$. This is
%closest in properties to {\bf model D}.

\section{Numerical Method for Modeling SSC Formation}

\begin{figure}[b]
\begin{center}
\includegraphics[width=.65\textwidth]{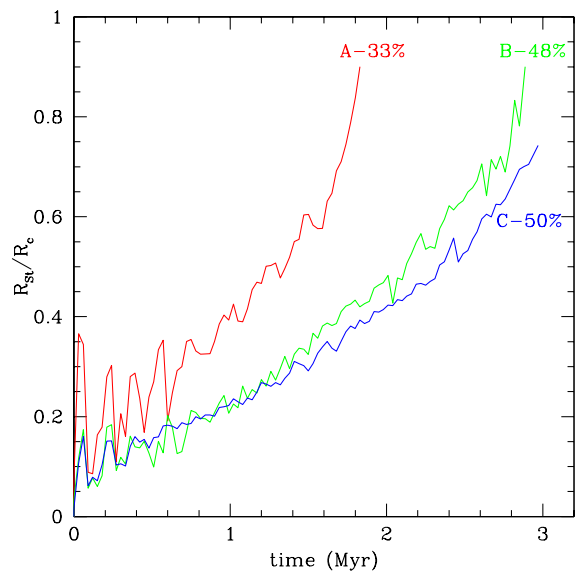}
\end{center}
\caption[]{Star cluster formation and cloud destruction at constant star formation rates, $\phi_{50}$. {\bf A}: $M_c=4000\:{\rm M_\odot}$ (Galactic case); {\bf B}: $M_c=4\times10^4\:{\rm M_\odot}$; {\bf C}: $M_c=4\times10^5\:{\rm M_\odot}$. Star formation efficiencies are quoted when $R_{\rm St}=0.9R_c$ or after 3 Myr.}

\label{fig:cldestnew}
\end{figure}

With an N-body code, we follow the dynamics, masses and sizes of a
collection of several thousand clumps, in the fixed potential of the
initial cloud and subject to the feedback processes described
above. At the same time we model the coupled evolution of a
spherically symmetric wind cavity and HII region. While we do not
account for clump collisions, we note that their initial mass spectrum
is probably created by a steady-state balance of collisional
fragmentation and agglomeration. Our model does yet include a physical
mechanism for star formation and so we investigate the response of our
fiducial gas clouds to different imposed steady star formation rates,
such that 50\% of the cloud would be converted to stars after
3~Myr. Stars are created from gas in the innermost clumps at these
rates and artificially added to the central cluster. We start our
models by instantly forming $\sim 250\:{\rm M_\odot}$ of stars,
equivalent to an ionizing luminosity of $S_{49}=1$. We examine the
time taken for the HII region to reach 90\% of the initial cloud
radius, at which point star formation is assumed to cease.
Photoionization gradually destroys our model clouds, {\bf A}, {\bf B}
and {\bf C}, as shown in figure \ref{fig:cldestnew}.

\section{Discussion}

The turbulent, clumpy and self-gravitating nature of our clouds
impedes their destruction and allows star formation to proceed to
higher efficiencies. For example, an HII region in a uniform,
quiescent cloud with the same mean density and initial star cluster as
model {\bf A}, would destroy the cloud in $\sim3\times 10^5$ years,
even with no additional star formation. However, with most of the
cloud mass in dense clumps with velocities set by virial equilibrium,
the process of ``turbulent mass loading'' confines the ionized gas for
1.5 -- 2 Myr, even though by this time, five times the initial stellar
mass has formed and feedback is correspondingly greater (model {\bf
A}). Reasonable models for clouds forming SSCs survive up to 3 Myr
(models {\bf B} and {\bf C}). The fraction of the HII region occupied
by the wind cavity increases with star cluster mass.

After a few Myr our neglect of Wolf-Rayet winds and supernovae becomes
important. Furthermore, our use of a fixed gravitational potential,
our assumption that all massive stars are at the cloud center and the
absence of protostellar winds in our model cause us to overestimate the cloud
destruction time.
%confines the HII region relative to the expansion expected into a
%quiescent, uniform medium of the same mean density. Our largest cloud,
%{\bf C}, with initial mass $4\times 10^5\:{\rm M_\odot}$ confines
%feedback for longer than 3 Myr, from a star cluster forming at a rate
%such that by this time 50\% of the initial mass is in stars. At this
%point Wolf-Rayet winds and supernovae, which we do not model, will
%accelerate the cloud destruction. Our use of a fixed gravitational
%potential, our assumption that all massive stars are at the center of
%the cloud and our neglect of protostellar winds further extend the
%cloud lifetime. 
Nevertheless, our results are qualitatively consistent with observed
ages of the youngest optically visible SSCs, which are $\sim 5\:{\rm
Myr}$ \cite{whi99}. Younger clusters are still embedded in dense gas
\cite{gil00,men00}. R136 in 30 Doradus, which is considered a small
SSC, has recently dispersed its gas and is $\sim 1-2$ Myr old
\cite{mas98}.

With steady star formation rates over the cloud lifetime, our results
imply that star formation efficiency increases with initial cloud
mass. The high efficiencies apparent in our SSC models allow for the
creation of bound clusters \cite{lad84}, even in the presence of
vigorous feedback. Since the mass loss is gradual, efficiencies as low
as $\sim30\%$ may result in loosely-bound clusters. Our Galactic model
({\bf A}) is close to this limit, though the inclusion of additional
feedback processes, such as protostellar winds, will reduce the
efficiency.

%The high star formation efficiencies apparent in our SSC models allow
%for the formation of bound stellar systems \cite{lad84}, even in the
%presence of vigorous feedback. In contrast, typical Galactic high mass
%star forming regions are not capable of such efficiencies. Our models
%will eventually be able to predict a minimum stellar mass required to
%form a bound cluster. Currently there are only a few measurements of
%SSC masses. Ho \& Filippenko \cite{ho96a,ho96b} measured stellar
%$\sigma_{\rm 1D}\simeq 11,16 \:{\rm km\:s^{-1}}$ for two SSCs in
%nearby dwarf galaxies, implying $M_*\simeq8,33\times 10^4\:{\rm
%M_{\odot}}$ and $\rho_{\rm *}\simeq2.7,1.1\times 10^4\:{\rm
%M_\odot\:pc^{-3}}$, for sizes of $0.9,1.9\:{\rm pc}$ and assumed
%virial equilibrium. However, Sternberg's \cite{ste98} reanalysis of
%these clusters found larger masses of $2.7\times 10^5\:{\rm M_\odot}$
%and $1.1\times 10^6\:{\rm M_\odot}$, respectively. Mengel et
%al. \cite{men00} find masses $\sim 0.5-5\times 10^6\:{\rm M_\odot}$
%for several IR-bright star clusters in the Antennae.

We plan to extend our models to include additional feedback processes,
such as protostellar winds, Wolf-Rayet winds and supernovae, and a
physical star formation mechanism (e.g., photoionization regulated
\cite{mck89}) in the neutral clumps. We shall compare our models to
observations of extra-galactic compact HII regions (e.g.,
\cite{kob99}).  We hope to predict minimum cloud masses and
densities required to form bound stellar systems and probe in more detail the
differences between Galactic and super star clusters.

\end{document}